# IEEE Copyright notice:

# Frequency Control of Decoupled Synchronous Machine Using Koopman Operator Based Model Predictive

Xiawen Li
Jaime De La Ree
Department of Electrical and Computer Engineering
Virginia Polytechnic Institute and State University
Blacksburg, U.S.
xiawenli@vt.edu

Chetan Mishra
Dominion Energy
Richmond, U.S.
Chetan.Mishra@dominionenergy.com

*Abstract*—Conventional generators have been retired or replaced by renewable energy because of the utility long-standing goals. However, instead of decommissioning the entire plant, the rotating mass can be utilized as a storage unit to mitigate the frequency issues due to these changes in the grid. The goal is to design a control utilizing the retired machine interfaced with the grid through a back to back converter referred to as decoupled synchronous machine system (DSMS) to damp frequency oscillations. However, in a practical setting, it is often not possible for a utility to obtain access to the detailed state equations of such devices from the vendor making the addition of another layer of control a challenging problem. Therefore, a purely data driven approach to nonlinear control design using Koopman operator based framework is proposed for this application. The effectiveness of the proposed system is demonstrated in the Kundur two-area system.

*Index Terms*—Frequency Control, Koopman operator, Model predictive control.

## I. INTRODUCTION

Owing to the recent EPA regulations [1], utilities are forced to retire more and more coal units. On top of that, the increasing penetration of renewable energy resources (RESs) is expected to introduce serious reliability concerns, frequency stability being one of the major ones. This is because unlike the conventional system, the RESs have very low or non-existent inertial response unless those functionalities for grid support are built in. That being said, currently, it is difficult to enforce those especially for distribution connected resources. This becomes a serious issue especially for systems with high penetration of solar PV generation since the yearly peaks are during light load months resulting in scenarios where a high percentage of the generation online is effectively inertia less making the frequency less robust to disturbances [2]. Although the variable speed wind turbines have a certain amount of inertia, they are usually decoupled from the network by power electronic devices unless additional controls are designed. Based on the current trends in the growth of such generation, this problem is expected to magnify [3].

In order to provide sufficient inertia, a lot of work has been done in fast frequency control techniques. References [4]–[6] use the power electronic converter with a suitable controller to emulate the inertia response of conventional power plants. The work in [7] runs solar PV at a lower output in order to have reserve power to participate in frequency regulation. The cooperation between the RESs and the energy storage systems (ESSs) is more commonly used to help damping frequency oscillation [8]. Our proposed approach utilizes the rotating mass of a decommissioned conventional generating unit as a storage interfacing it to the grid via a back to back converter configuration for fast control.

Wide area control in large scale systems is usually not an option due to the cost prohibitive communication infrastructure requirements which leaves us with the option of distributed control. Compared to wide area control, distributed algorithms need only limited amounts of information shared between agents which therefore improve cybersecurity and superior computation speed. Furthermore, distributed control has higher robustness regarding the failure of other agents [9].

Nearly every control application imposes constraints on the input controls and the states. The existence of hard constraints results in the need of control strategy for dealing with them. Normally, determining the feedback solution for the cost function satisfying those constraints cannot be found analytically because it involves the solution of the Hamilton Jacobi-Bellman differential or difference equation which leads to an extremely more difficult task [10]-[11]. Model predictive control (MPC) is one of few suitable approaches, with the fact that this method circumvents the close solution by the repeated solution of a finite-horizon open-loop optimization problem using the current states of the system as the initial states. The effectiveness of MPC is shown in [12]–[14]. Power systems are non-linear [15] thus requiring models for predicting the evolution of states which is called nonlinear MPC [11]. However, this approach is computationally prohibitive and therefore a lot of work has been done in using linear MPC for controlling nonlinear dynamical systems [16], [17]. Thus, an approach for linearization in the large region is needed to improve the accuracy of prediction for a nonlinear system in order to use a linear MPC in regions further from the equilibrium.

That being said, due to the inherently nonlinear nature of the power system and uncertain characteristics of the converter-interfaced generation, an accurate explicit model of the power grid is of the great challenge making a data driven control the only practical option. The Koopman operator theory is a popular and powerful tool which is capable of the analysis and prediction of the nonlinear dynamical system using measurement data [18] making it a data driven approach. The Koopman operator utilizes a set of scalable observable functions to reconstruct the underlying dynamical system from measurement data in a linear but high-dimension space [18] i.e. the system is linear in this new space which makes it an attractive option to be used with linear MPC. Building upon the recent development of the Koopman model predictive control (KMPC) framework [19], a feedback control of the local frequency of the DSMS is proposed. KMPC has been found to be an effective solution to this problem in [18]–[21].

The remainder of the paper is organized as follows. Section II presents the dynamic model of the proposed system and the problem statement. Section III describes the fundamental idea behind the KMPC. The proposed system is demonstrated in Kundur two-area system in Section IV. Finally, Section V concludes the paper.

## II. SYSTEM MODEL AND PROBLEM FORMULATION

### A. System Model

#### 1) Dynamics of the decoupled synchronous machine

The dq0 frame is used in the derivation of models of electrical machines in this paper since it is well suited for electro-mechanical transient studies. It also gives a suitable representation of the voltage source converter (VSC) averaged model. The q-axis of the rotating frame is aligned with the AC-side voltage. The decoupled synchronous machine is represented by the classical model which is a constant voltage source $E_q$ behind the transient reactor $R_s + jX_s$. The system schematic is shown in Fig.1. The dynamics of the system is given by the following equations.

$$\begin{aligned}
\frac{X_s}{\omega_r}\frac{di_{sd}}{dt} &= X_s i_{sq} - R_s i_{sd} - v_{sc}^d \\
\frac{X_s}{\omega_r}\frac{di_{sq}}{dt} &= -X_s i_{sd} - R_s i_{sq} - v_{sc}^q + E_q \\
\frac{d\delta_r}{dt} &= \omega_r \\
\frac{2H}{\omega_0}\frac{d\omega_r}{dt} &= -P_e \\
\frac{X_g}{\omega_s}\frac{di_{gd}}{dt} &= X_g i_{gq} - R_g i_{gd} - v_{gc}^d \\
\frac{X_g}{\omega_s}\frac{di_{gq}}{dt} &= -X_g i_{gd} - R_g i_{gq} - v_{gc}^q + V_{gq} \\
\frac{dv_{dc}}{dt} &= \frac{1}{C_{dc}}(i_{dc} - i_L)
\end{aligned} \quad (1)$$

where $i_{kdq}$ is dq-axis current. $R_g + jX_g$ is the grid-side converter phase reactor. $v_{kc}^{dq}$ is the dq-axis converter voltage. $\omega_r$ is the electrical rotor speed deviation relative to the nominal frequency $\omega_0$ in radians per second [rad/s]. $\omega_s$ is the grid frequency. $V_{gq}$ is the q-axis grid voltage. The DC circuit is modeled by the equivalent DC capacitance $C_{dc}$, $v_{dc}$, $i_{dc}$ and $i_L$. k = s represents the machine-side parameter and k = g the grid-side parameter. The power balance between the AC and DC sides is given as

$$P_e = E_q i_{sq} = v_{dc}i_{dc} = v_{sc}^q i_{sq} + v_{sc}^d i_{sd} = v_{gc}^q i_{gq} + v_{gc}^d i_{gd} = V_{gq}i_{gq} \quad (2)$$

where $P_e$ is the electrical output power provided by this proposed system.

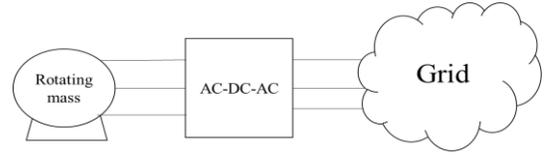

Figure 1. DSMS system diagram

#### 2) Converter Controller Structure

The goal of the converter controller control is mainly to control the active and reactive power at the AC grid terminal. In this paper, the reactive power reference is set to zero since the objective is to damp frequency oscillation by absorbing or delivering active power. The overall configuration of the controller of DSMS is shown in Fig. 2.

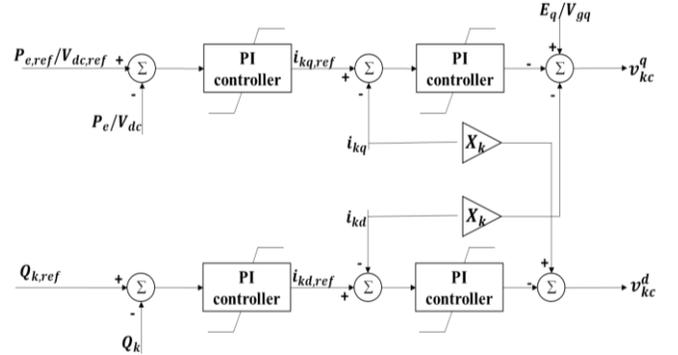

Figure 2. DSMS control structure

### B. Problem Formulation

The goal of this paper is to mitigate the frequency oscillation by adjusting the active power output $P_e$ of the proposed DSMS using linear KMPC. As introduced in section I, distributed control has the advantages with respect to data privacy, constraints, etc., and it is therefore adopted in this paper.

## III. KOOPMAN MODEL PREDICTIVE CONTROL

### A. Koopman Operator Theory

Consider a discrete-time nonlinear uncontrolled dynamical system

$$x_{k+1} = f(x_k) \tag{3}$$

where $x_k \in M \subset \mathbb{R}^n$ is the state vector of the system at time step $k$. The Koopman operator $\mathcal{K}$ is defined in the following form

$$\mathcal{K}g(x) = g(f(x)) \tag{4}$$

where $g : \mathbb{R}^n \to \mathbb{R}$ is a set of scalar valued observable functions which forms the infinite-dimensional space $M$. The Koopman operator $\mathcal{K}$ is a linear representation of the nonlinear dynamical system on space $M$. This operator significantly captures dynamics of the underlying system in the space of observables $M$ which consists of the components of $x$.

The nonlinear controlled dynamical system is given by

$$x_{k+1} = f(x_k, u_k) = y \tag{5}$$

where $u_k \in U \subset \mathbb{R}^m$ is the control input of the system and $y$ is the output of the system. The corresponding Koopman operator is therefore required to work on the extended state space which is defined as the Cartesian product $M \times l(U)$ of the original state space and the space of all input sequences [18]. The system with extended state is denoted as

$$\chi_{k+1} = F(\chi_k) = \begin{bmatrix} f(x_k, u_k(0)) \\ Su_k \end{bmatrix} \tag{6}$$

where

$$\chi = \begin{bmatrix} x_k \\ u_k \end{bmatrix} \tag{7}$$

$$Su_k(i) = u_{k+1}(i+1) \tag{8}$$

where i denotes the i$^{th}$ element in the input sequences.

### B. Construction of Koopman Operator based Linear System

This section introduces the method for constructing Koopman operator based linear system proposed in [19]. Based on the previous discussion, with a correct state transformation, the system can be represented by a linear system in the new state space. We assume that the states are not directly measured and we only have access to the input $u$ and output measurements $y$ where $y$ is the frequency on the grid side and $u$ is the real power output from the DSMS. Therefore, we look for new state z as a non-linear function of y resulting in a linear time invariant system of the form,

$$z_{k+1} = Az_k + Bu, \quad z \in R^N, u \in R^m \tag{9}$$

where $N \gg n$. Let us define the observable functions $g$ as a set of nonlinear functions that map the output $y$ to z i.e.

$$g(y) = [g_1(y), ..., g_N(y)]^T, \quad z = g(y) \tag{10}$$

Now a correct choice of $g$ is extremely important to have a truly linear system in a large region in the original state space. A single frequency measurement is insufficient to observe the system dynamics and therefore it alone is not a good choice for $g$. Therefore, we first take high order derivatives of the frequency and inputs themselves in the form of time-delay embedding $\zeta$ given in (11) [18], [19].

$$\begin{aligned} \zeta_i &= \left[ y_i^T, ..., y_{i+nd-1}^T, u_i^T, ..., u_{i+nd-1}^T \right]^T \\ \zeta_i^+ &= \left[ y_{i+1}^T, ..., y_{i+nd}^T, u_{i+1}^T, ..., u_{i+nd}^T \right]^T \end{aligned} \tag{11}$$

where $i = j, ..., j + nd - 1$ at some $j$ time stamp and $nd$ is the number of delays.

The vector $g$ in (10) is then defined as

$$g = \left[ \zeta_i, \|\zeta_i(1)\|_2, constant \right]^T \tag{12}$$

The constant term is introduced as an element of z to allow for a non zero vector field when other z's are 0. The system matrices A and B are then obtained through linear regression as follows.

$$\min_{A,B} \|Z_{k+1} - AZ_k - BU\|_F \tag{13}$$

$$[A \quad B] = Z_{k+1}[Z_k \quad U]^\dagger \tag{14}$$

### C. Model Predictive Control

In this section, the KMPC formulation is provided. This predictor given in (9) allows the application of the well-known linear MPC tools to control the nonlinear system. Furthermore, the computation of KMPC is very fast allowing for online application since the dimension of the embedded space does not impose impact on the computational complexity of MPC. The KMPC controller is solved at every time step of the following optimization problem (see [19] for more details). The objective function is formulated to obtain the optimal input $u=P_e$ to achieve the desired output $z^{ref}= \omega_s^{ref}$.

$$\begin{aligned} \min_{u_i, y_i} &\sum_{i=0}^{N_p-1} (z_i^{ref} - z_i)^T Q(z_i^{ref} - z_i) + u_i^T R u_i \\ s.t. \quad &z_{i+1} = Az_i + Bu_i, \quad i = 0, ..., N_p \\ &|P_{e,i}| \leq b_i, \quad i = 0, ..., N_p - 1 \\ &z_0 = g(\zeta_0) \end{aligned} \tag{15}$$

where $Q$ is the symmetric positive semidefinite weighting matrix. The structure of $Q$ is chosen such that it only tries to minimize the norm of frequency. $R$ is set to 1. $\zeta_0$ denotes the most recent value for the vector $\zeta$. The whole optimization process is repeated in a receding horizon fashion where only the first element of the input sequence $u$ is applied at each instance.

## IV. NUMERICAL RESULTS

In this section, the proposed system shown in Fig. 3 is illustrated through the *Simulink* sample case Kundur two-area system. The Generator No.3 is replaced with the DSMS system which consists of a 900 MW base power, 20 kV rated voltage machine. It has a 0.0025+j0.25 pu external impedance and 6.175s inertia coefficient. The data of the grid-side inverter is obtained based on [12]. The DC-link voltage is

20√3 kV with 0.004 pu capacitor. The phase impedance of the inverter is 0.001 + j0.1656 pu.

In the present work, we design a control for a given post fault system. In order to design the KMPC constructed in the form (15), data from $10^4$ trajectories starting from different initial points in state space with 50ms sampling time was collected using time domain simulations done offline. The initial conditions of the states are drawn uniformly from the intervals; $\Delta\omega_i \in [-0.05, 0.05]$ and $\Delta\delta_i \in [-20°, 20°]$ where $\Delta\delta_i$ and $\Delta\omega_i$ are the rotor angle deviations from the initial steady-state values and the rotor speed deviations from the nominal frequency $\omega_b$ of the i[th] synchronous generator. The dynamics of DSMS is not counted into the states because it is much faster than the dynamics of the external system. The control input $P_e$ is sampled from $[-0.3, 0.3]$. The values shown are in per unit. The prediction horizon for KMPC is chosen to be 0.5s which yields $N_p$=10. The quadratic program (15) is solved using qpOASES [22].

To demonstrate the effectiveness of the approximation of this system in the lifted space, we compare the predictions with true measurements starting from a random initial condition $[\Delta\delta_i, \Delta\omega_i] = [-3.88, 5.6, -11.84, 0.0179, 0.0258, 0.0243]^T$ under the same control sequences in Fig. 5. The relative root mean squared errors (RMSE) over 100 randomly sampled initial conditions was calculated to be 0.1468 using the formula below.

$$RMSE = 100 \cdot \frac{\sqrt{\sum_k \|x_{pred}(kT_s) - x_{true}(kT_s)\|_2^2}}{\sqrt{\sum_k \|x_{true}(kT_s)\|_2^2}} \quad (16)$$

Now, we compare the performance of the proposed control by comparing the frequency response for three systems under same set of faults. System A is the original system. System B is the original system with generator No.3 replaced by constant negative load in order to mimic an inertia less renewable generation without grid support functionality. Lastly, System C is the system with generator No.3 replaced by the constant negative load and the proposed DSMS.

The fault No.1 being applied one is a three-phase-to-ground fault at bus 7 at 2s. The line between bus 7 and 8 is tripped after 6 cycles and reclosed at 5s. The frequency response of the three systems is given in Fig. 6. It can be seen from the plots that on displacing the conventional generator in system A with inertia less negative load, the frequency swings became bigger as expected. That being said, in System C, the frequency excursions are smaller with much faster damping in the oscillations. Furthermore, one can also see the benefit of adding the converter interface to the rotating mass as this combined system performs much better than the original system A with the synchronous machine having the same inertia.

Now, to check the robustness of the proposed control for a different fault resulting in a different post fault topology, we study a fault at bus 9 resulting in a post fault system with the line 7-9 tripped to clear the fault and reclosed at 5s. It can be seen in Fig. 7 that the proposed control system performs decent enough. That being said, change of system parameters would definitely require a choice of more optimal Koopman operator for the same set of observable functions which will be explored in future work.

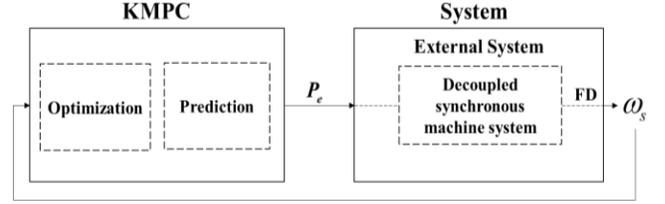

Figure 3. Proposed system structure

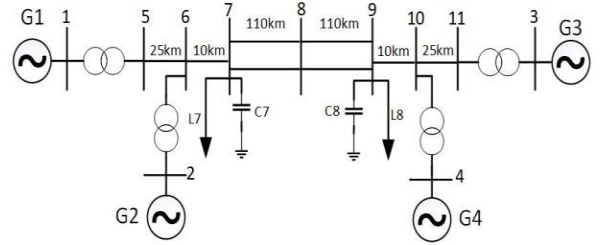

Figure 4. Kundur two-area sytem

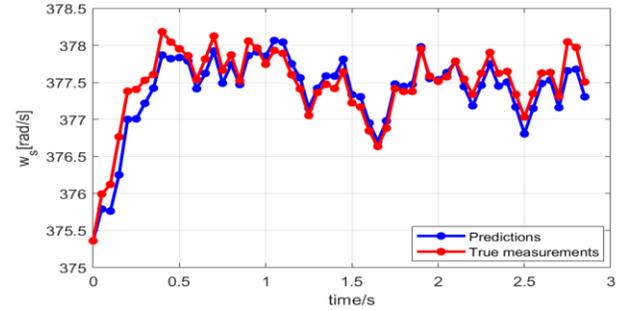

Figure 5. Predicted frequency V.S. Truely measured frequency

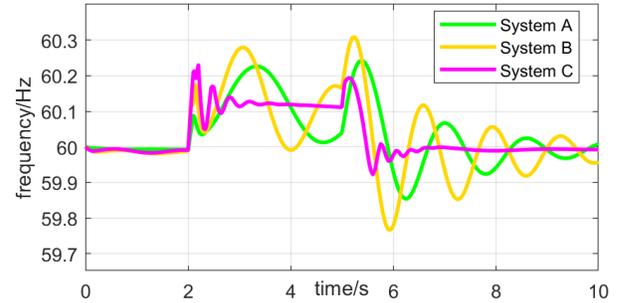

(a) Frequency responses

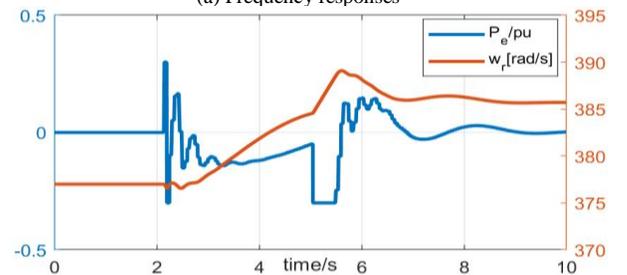

(b) Control input & rotor speed of DSMS

Figure 6. Responses of fault No.1

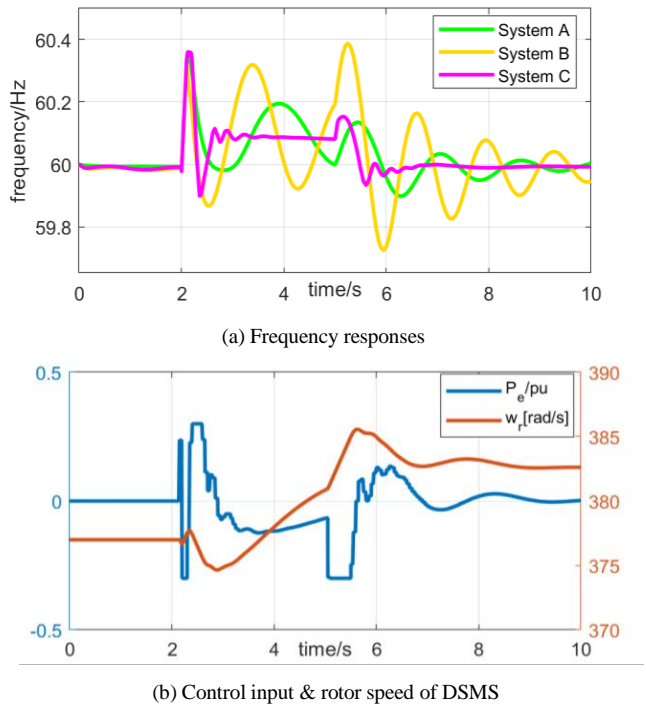

(a) Frequency responses

(b) Control input & rotor speed of DSMS

Figure 7. Response of fault No.2

## V. CONCLUSION

A KMPC based DSMS is proposed in this paper. The DSMS consists of the converter-based system and a retired synchronous generator with no governor used as a storage. The overall approach involves estimating the state matrices for a linear system in a transformed nonlinear state space using offline time domain simulations for a given post-fault system. The resulting system showed very promising results when estimating the trajectory for the original nonlinear system. Next, a linear KMPC is proposed in the transformed state space to damp frequency oscillations under large disturbances like three-phase faults. The effectiveness of the proposed controller is demonstrated in the Kundur two-area system. The proposed control also shows a decent performance for a fault resulting in a totally different post fault system giving an idea of robustness to changes in system parameters. That being said, there is still a lot to be explored in this direction and therefore, adaptive control in Koopman operator framework will be attempted in future.